\documentstyle[aps,prd]{revtex}
\begin{document}
\hfill{hep-ph/0102013}\par
\vskip 0.3cm
\centerline{\large \bf Threshold resummation for exclusive
$B$ meson decays}
\vskip 0.3cm
\centerline{Hsiang-nan Li\footnote{E-mail: hnli@phys.sinica.edu.tw}}\par
\centerline{Institute of Physics, Academia Sinica,
Taipei, Taiwan 115, Republic of China}
\centerline{Department of Physics, National Cheng-Kung University, 
Tainan, Taiwan 701, Republic of China}\par
\vskip 0.5cm

\par
\vskip 0.5cm

\centerline{\bf Abstract}
\vskip 0.3cm

We argue that double logarithmic corrections $\alpha_s\ln^2 x$ 
need to be resumed in perturbative QCD factorization theorem for 
exclusive $B$ meson decays, when the end-point region with a momentum 
fraction $x\to 0$ is important. These double logarithms, being of the 
collinear origin, are absorbed into a quark jet function, which is 
defined by a matrix element of a quark field attached by a Wilson 
line. The factorization of the jet function from the decay 
$B\to\gamma l\bar\nu$ is proved to all orders. Threshold resummation 
for the jet function leads to a universal, {\it i.e.}, 
process-independent, Sudakov factor, whose qualitative behavior is 
analyzed and found to smear the end-point singularities in heavy-to-light 
transition form factors.

\section{INTRODUCTION}

Perturbative QCD (PQCD) factorization theorem for exclusive processes
\cite{BL} states that a hadronic form factor at large momentum transfer
is expressed as convolution of a hard amplitude with hadron distribution
amplitudes. However, the PQCD evaluation of the pion form factor suffers
the soft enhancement from the end point of a momentum fraction $x\to 0$
\cite{IL}. In this region the hard amplitude is characterized by a low
scale, such that perturbative expansion in terms of a large coupling
constant $\alpha_s$ is not self-consistent. More serious end-point
singularities, logarithmic and linear, have been observed in the 
leading-twist (twist-2) and next-to-leading-twist (twist-3) 
contributions to the $B\to\pi$ transition form factors
\cite{SHB,ASY,BF}, respectively.

We argue that when the end-point region is important, the
double logarithms $\alpha_s\ln^2 x$ from radiative corrections 
should be organized to all orders in order to improve perturbative
expansion. These double logarithms have been found in the semileptonic
decay $B\to\pi l\bar\nu$ \cite{ASY} and in the radiative
decay $B\to\gamma l\bar\nu$ \cite{KPY}, and resummed. Here
we shall give a systematic treatment of this type of double logarithms,
which is of the collinear origin, by introducing a quark jet function 
into PQCD factorization theorem. The all-order factorization of the 
jet function from the decay $B\to\gamma l\bar\nu$ is proved 
following the procedures proposed in \cite{L4,NL}, which provides
a solid theoretical ground for the modified formalism appropriate for the
end-point region. It will be shown that the jet function is
defined as a matrix element of a quark field attached by a Wilson line.

The double logarithms in the jet function are resummed into a Sudakov 
form factor in the Mellin space using the technique developed in 
\cite{S0,CT,L1,L2}. In principle, with the definition of the jet function 
constructed in this work, its threshold resummation can be performed up 
to the next-to-leading-logarithm accuracy. Since our purpose is to 
understand the qualitative behavior of the jet function at small $x$, it 
suffices to derive only the leading-logarithm result. 
Extending the above formalism to the semileptonic decay 
$B\to\pi l\bar\nu$, we obtain the identical Sudakov factor, implying its 
universality. By means of the inverse Mellin transformation, the Sudakov 
factor is found to vanish quickly as $x\to 0$. The suppression is so 
strong that the end-point singularities in the $B\to\pi$ form factors 
are smeared. We conclude that in a self-consistent 
perturbative evaluation of heavy-to-light transition form factors, the 
end-point singularities do not exist.

It has been argued that the dependence on
the transverse momentum $k_T$ carried by the light spectator in a
$B$ meson can not be dropped from hard amplitudes for exclusive $B$ 
meson decays \cite{KPY}. As considering higher-order corrections to a
hard amplitude, $k_T$ appears in the ratio $k_T/k^+$, $k^+$
being the longitudinal component of the light spectator momentum.
Obviously, this ratio is not power-like, because $k_T$ and $k^+$ are
of the same order of magnitude. Hence, a reliable formalism for 
exclusive $B$ meson decays must involve the parton transverse degrees 
of freedom. The effect due to the inclusion of $k_T$ has been explored 
thoroughly in \cite{LY1,KLS,LUY}, which also smears the end-point 
singularities in the heavy-to-light form factors \cite{TLS}. In this 
paper we shall concentrate only on the Sudakov effect from threshold 
resummation, and neglect the $k_T$ dependence.
    
In Sec.~II we work out the $O(\alpha_s)$ factorization formula for the 
$B\to\gamma l\bar\nu$ decay amplitude at the end point. The all-order 
factorization is proved in Sec.~III, leading to the convolution of the
hard amplitude with the $B$ meson distribution amplitudes and with the jet 
function. Threshold resummation is done in Sec.~IV by solving an
evolution equation for the jet function. 
In Sec.~V we extract the jet function from the 
semileptonic decay $B\to\pi l\bar\nu$. The smearing effect from the 
Sudakov factor on the end-point singularity is demonstrated 
numerically by evaluating the leading-twist contribution to the 
$B\to\pi$ form factor. Section VI is the conclusion. In Appendix A 
we comment on the Sudakov resummation performed in \cite{ASY}.
  
\section{$O(\alpha_s)$ FACTORIZATION}

We study the radiative decay $B\to\gamma l\bar\nu$. The $B$ meson 
carries the momentum $P_1=(M_B/\sqrt{2})(1,1,{\bf 0}_T)$ and the outgoing 
photon carries the momentum $P_2=(M_B/\sqrt{2})(0,\eta,{\bf 0}_T)$ with 
the energy fraction $\eta$, where the light-cone variables have been 
adopted. Consider the kinematic region with small 
$q^2=(1-\eta)M_B^2$, $q=P_1-P_2$ being the lepton pair momentum, 
{\it i.e.}, with large $\eta$, where PQCD is applicable. The mass 
difference between the $B$ meson and the $b$ quark, $\bar\Lambda=M_B-m_b$, 
is treated as a small scale in the heavy quark limit. As stated in \cite{L4}, 
there are two types of infrared divergences in radiative corrections, 
soft and collinear. Soft divergences come from the region of a loop 
momentum $l$, where all its components vanish. Collinear divergences are 
associated with a massless parton carrying a momentum of order $M_B$.
The $O(\alpha_s)$ factorization of the soft divergences into the 
$B$ meson distribution amplitudes has been presented in \cite{L4}. 
Here we derive only the $O(\alpha_s)$ factorization of the collinear
divergences at the end point into the jet function.

According to leading-twist PQCD factorization theorem, the
form factor relevent to the decay $B\to\gamma l\bar\nu$ is written as 
\begin{eqnarray}
G(q^2)=\sum_{m=+,-}\phi_m(x)\otimes H_m(x,\eta)\;,
\label{g2}
\end{eqnarray}
where the symble $\otimes$ denotes the convolution over the
spectator momentum fraction $x=k^+/P_1^+$, $k$ being the spectator quark 
momentum. The light-cone $B$ meson distribution amplitudes $\phi_m$ are 
defined by \cite{L4,NL} 
\begin{eqnarray}
\phi_{\pm}(x)&=&\int \frac{dy^-}{2\pi}e^{ixP_1^+y^-}
\langle 0|{\bar q}(y^-)\gamma_5(\not v+I)\gamma^{\pm}
\exp\left[-ig\int_0^{y^-}dzn_-\cdot A(zn_-)\right]b_v(0)
|B(P_1)\rangle\;,
\label{b1}
\end{eqnarray}
with the dimensionless vectors $v=P_1/M_B$ and $n_-=(0,1,{\bf 0}_T)$, and the
rescaled $b$ quark field $b_v$. The hard amplitudes $H_m$ are obtained by 
contracting the quark-level diagrams with the spin structures corresponding 
to $\phi_m$ \cite{GN}. Figure 1 is the lowest-order example, where the upper 
line represents a $b$ quark and $\times$ represents a weak decay vertex. The 
contributions to the hard amplitudes from Figs.~1(a) and 1(b) scale 
like $1/({\bar\Lambda} M_B)$ and $1/M_B^2$, respectively. Below 
we shall concentrate on Fig.~1(a).

The factorization formula in Eq.~(\ref{g2}) is appropriate for the region 
with $k^+\sim O(\bar\Lambda)$, in which the only infrared divergences are
the soft ones absorbed into $\phi_\pm$ in Eq.~(\ref{b1}). Near the end point 
$k^+\sim O(\bar\Lambda^2/M_B)$, the internal quark in Fig.~1(a) carries a 
large momentum $P_2-k$ with its invariant mass vanishing like 
$(P_2-k)^2=-2xP_1\cdot P_2\sim O(\bar\Lambda^2)$. This kinematics is 
similar to the threshold region of deeply inelastic scattering (DIS) with the 
Bjorken variable $x_B\to 1$, where the scattered quark also carries a 
momentum with a large component and possesses a small invariant mass 
$(1-x_B)s$, $s$ being the center-of-mass energy. In this region the 
scattered quark produces a jet of particles, to which the radiative corrections 
contain additional collinear divergences. Hence, a jet function needs to be 
introduced into the standard factorization theorem for DIS \cite{G}. 
Similarly, a jet function has been incorporated into the factorization of 
direct photon production at a large photon transverse momentum 
(threshold) \cite{LSV}. Here we associate a jet function 
with the internal quark at the end point of the momentum
fraction involved in the decay $B\to\gamma l\bar\nu$.

An additional collinear divergence from the loop momentum parallel to 
$P_2$ appears in the higher-order correction to 
the weak decay vertex shown in Fig.~1(c). This divergence can be 
extracted by replacing the $b$ quark line by an eikonal line in the 
direction of $n_+$:
\begin{eqnarray}
\frac{\not P_1-\not k+\not l+m_b}{(P_1- k+ l)^2-m_b^2}\gamma^\beta
b(P_1-k)\approx \frac{n_+^\beta}{n_+\cdot l} b(P_1-k)\;,
\label{ib}
\end{eqnarray}
where $ b(P_1-k)$ represents the $b$ quark spinor with the momentum 
$P_1-k$ and $n_+=(1,0,{\bf 0}_T)$ is a dimensionless vector. 
The above replacement makes sense, since $\gamma^\beta$ is 
dominated by its plus component $\gamma^+$, $l^-$ is much larger 
than $l^+$, and $\not l\gamma^\beta$ diminishes. 
The factorization of the fermion flow is achieved by
inserting the Fierz identity,
\begin{eqnarray}
I_{ij}I_{lk}=\frac{1}{4}I_{ik}I_{lj}
+\frac{1}{4}(\gamma_5)_{ik}(\gamma_5)_{lj}
+\frac{1}{4}(\gamma_\alpha)_{ik}(\gamma^\alpha)_{lj}
+\frac{1}{4}(\gamma_5\gamma_\alpha)_{ik}(\gamma^\alpha\gamma_5)_{lj}
+\frac{1}{8}(\sigma_{\alpha\beta})_{ik}(\sigma^{\alpha\beta})_{lj}\;,
\label{fi}
\end{eqnarray}
in which the first and last terms contribute in the combined
structure,
\begin{eqnarray}
I_{ij}I_{lk}\to \frac{1}{4}I_{ik}(\not n_+\not n_-)_{lj}\;.
\label{fi1}
\end{eqnarray}

Assigning the identity matrix $I$ in Eq.~(\ref{fi1}) to the trace for the hard 
amplitude, we obtain Fig.~1(a). The matrix $\not n_+\not n_-/4$ then leads to 
the loop integral, 
\begin{eqnarray}
J_{\parallel}^{(1)}&=&-ig^2C_F\int\frac{d^4 l}{(2\pi)^4}
\frac{1}{4}tr\left[\not n_+\not n_-
\gamma_\beta\frac{\not P_2-\not k+\not l}
{(P_2-k+l)^2}\right]\frac{n_+^\beta}{n_+\cdot l l^2}\;,
\nonumber\\
&=&-\frac{\alpha_s}{4\pi}C_F\ln^2 x+\cdots\;,
\label{j1c}
\end{eqnarray}
where $C_F=4/3$ is a color factor. To obtain the above double logarithm, we 
have regularized the collinear pole from $l$ parallel to $n_+$ by allowing 
$n_+$ to possess a small amount of minus component. The correction to the 
photon vertex in Fig.~1(d) contains only the single logarithm $\alpha_s\ln x$, 
since the phase space of the loop momentum is restricted to 
$0<l^+<k^+\sim O(\bar\Lambda^2/M_B)$. The self-energy correction in 
Fig.~1(e), generating also the single logarithm, is factorized into 
$J_{\perp}^{(1)}$ trivially by applying the Fierz transformation in 
Eq.~(\ref{fi1}). The explicit expression of $J_{\perp}^{(1)}$ is not essential 
here. The reason for labeling the $O(\alpha_s)$ jet functions
by $\parallel$ and $\perp$ will become clear in the next section.
To group Figs.~1(c) and 1(e), we introduce a quark jet function with 
the zeroth-order expression and the $O(\alpha_s)$ expression,
\begin{eqnarray} 
J^{(0)}=tr(\not n_+\not n_-)/4=1\;,\;\;\;\;
J^{(1)}=J_{\parallel}^{(1)}+J_{\perp}^{(1)}\;,
\end{eqnarray}
respectively. 

It has been shown that the soft divergences in Figs.~1(c), 1(d) and 
1(f)-1(h) are absorbed into the $O(\alpha_s)$ $B$ meson distribution 
amplitudes $\phi_\pm^{(1)}$ \cite{L4}. The remaining infrared 
finite $O(\alpha_s)$ contributions, including the single logarithms 
except that from Fig.~1(e), are assigned into the hard amplitudes 
$H_\pm^{(1)}$. Therefore, the modified factorization in the end-point 
region is written, up to $O(\alpha_s)$, as
\begin{eqnarray}
G^{(0)}+G^{(1)}=\sum_{m=+,-}[1+\phi_m^{(1)}]\otimes
[H_m^{(0)}+H_m^{(1)}]\otimes [1+J^{(1)}]+O(\alpha_s^2)\;,
\end{eqnarray}
with $G^{(0)}$ and $G^{(1)}$  displayed in Fig.~1. Note that $J$ does 
not depend on the subscript $m$, a fact obvious from the derivation 
of Eq.~(\ref{j1c}).

\section{ALL-ORDER FACTORIZATION}

In this section we prove the factorization theorem for the radiative 
decay $B\to \gamma l\bar\nu$ at the end point to all orders, and
construct the definition of the jet function
$J(x)$ via
\begin{eqnarray}
J(x)\bar u(P_2-k)\equiv
\langle u(P_2-k)|{\bar q}(0)\frac{1}{4}\not n_+\not n_-
\exp\left[-ig\int_{-\infty}^0dzn_+ \cdot A(zn_+)\right]|0\rangle\;.
\label{pwj}
\end{eqnarray}
The spinor $u(P_2-k)$ is associated with the internal quark, through 
which the momeutm $P_2-k$ flows. The idea of the proof
is based on induction \cite{L4,NL} with the help of the Ward identity,
\begin{eqnarray}
l_\mu G^\mu(l,k_1,k_2,\cdots, k_p)=0\;,
\label{war}
\end{eqnarray}
where $G^\mu$ is a physical amplitude with an external gluon
carrying the momentum $l$ and with $p$ external quarks carrying the
momenta $k_1$, $k_2$, $\cdots$, $k_p$. All these external particles are
on mass shell. 

The $O(\alpha_s)$ factorization of the collinear divergences has been 
worked out in Sec.~II. Assume that the factorization theorem holds up to 
$O(\alpha_s^N)$:
\begin{eqnarray}
G=\sum_{m=+,-}\phi_m\otimes H_{m}\otimes J\;,
\label{gh1}
\end{eqnarray}
with the definitions,
\begin{eqnarray}
G=\sum_{i=0}^N G^{(i)}\;,\;\;\;
\phi_{m}=\sum_{i=0}^N \phi_{m}^{(i)}\;,\;\;\;
H_{m}=\sum_{i=0}^N H_{m}^{(i)}\;, \;\;\;
J=\sum_{i=0}^N J^{(i)}\;, 
\label{ghf}
\end{eqnarray}
$G^{(i)}$ denotes the full diagrams of $O(\alpha_s^i)$. The $B$ meson 
distribution 
amplitudes $\phi_{m}^{(i)}$ and the jet function $J^{(i)}$ are defined 
by the $O(\alpha_s^i)$ terms in the perturbative expansions of 
Eq.~(\ref{b1}) and of Eq.~(\ref{pwj}), respectively. The $O(\alpha_s)$
hard amplitudes $H_{m}^{(i)}$ do not contain the intial-state soft  
divergences and the end-point double logarithms, which 
have been collected into the $B$ meson distribution amplitudes 
and into the jet function, respectively. 
We then have the relations,
\begin{eqnarray}
G^{(k)}=\sum_{m=+,-}\sum_{i=0}^{k}\sum_{j=0}^{k-i}
\phi_{m}^{(i)}\otimes
H_{m}^{(k-i-j)}\otimes J^{(j)}\;,\;\;\;k=0,1,\cdots N\;.
\label{gnf}
\end{eqnarray}

Below we prove the collinear factorization of the $O(\alpha_s^{N+1})$
diagrams $G^{(N+1)}$ into the convolution of the $O(\alpha_s^{N})$
diagrams $G^{(N)}$ with the $O(\alpha_s)$ jet function $J^{(1)}$.
Look for the gluon in a complete set of $O(\alpha_s^{N+1})$
diagrams $G^{(N+1)}$, one of whose ends attaches the lower most vertex 
on the internal quark line. Such a gluon exists, because $G^{(N+1)}$ are 
the finite-order diagrams. Let $\alpha$ denote the lower most
vertex, and $\beta$ denote the attachments of the other end of the
identified gluon inside the rest of the diagrams. There are two types of
collinear configurations associated with this gluon, depending on whether
the vertex $\beta$ is located on an internal line with a momentum along
$P_2$. The fermion propagator adjacent to the vertex $\alpha$ is
proportional to $\not P_2$ in the collinear region with the loop momentum
$l$ parallel to $P_2$. If $\beta$ is not located on a collinear line
along $P_2$, the component $\gamma^-$ in $\gamma^\alpha$ and the plus
component of the vertex $\beta$ give the leading contribution. If $\beta$
is located on a collinear line along $P_2$, $\beta$ can not be plus, and
both $\alpha$ and $\beta$ label the transverse components. This
configuration is the same as of the self-energy correction to an on-shell
particle.

According to the above classification, we decompose the tensor
$g_{\alpha\beta}$ appearing in the propagator of the identified gluon 
into
\begin{eqnarray}
g_{\alpha\beta}=\frac{n_{+\alpha} l_\beta}{n_+\cdot l}
-\delta_{\alpha \perp}\delta_{\beta \perp}
+\left(g_{\alpha\beta}-\frac{n_{+\alpha} l_\beta}{n_+\cdot l}
+\delta_{\alpha \perp}\delta_{\beta \perp}\right)\;.
\label{dec}
\end{eqnarray}
The first term on the right-hand side of Eq.~(\ref{dec}) extracts the first type 
of collinear divergences, since the light-like vector $n_{+\alpha}$ selects 
the minus component of $\gamma^\alpha$, and $l_\beta$ in the collinear 
region selects the plus component of the vertex $\beta$ . The second term
extracts the second type of collinear divergences. The last term does not 
contribute a collinear divergence. We shall concentrate on the factorization 
corresponding to the first term, and the factorization corresponding to the 
second term can be achieved simply by requiring gauge invariance 
\cite{NL}. 

The identified collinear gluon with $\alpha=-$ and $\beta=+$ does not 
attach the internal quark line directly, which carries a momentum along 
$P_2$ at the end point. That is, those diagrams with Fig.~1(e) as the 
$O(\alpha_s)$ subdiagram are excluded from the set of $G^{(N+1)}$ as 
discussing the first type of collinear configurations. Applying the Fierz 
transformation in Eq.~(\ref{fi1}) to break the fermion flow, we have the 
physical amplitude, in which the two on-shell quarks and the on-shell 
gluon carry the momenta $\xi P_2$, $\xi P_2-l$ and $l$, respectively. The 
fraction $\xi$ reflects that the momentum flowing through the internal quark 
is almost parallel to $P_2$. Figure 2(a), describing the Ward identity, 
contains a complete set of contractions of $l_\beta$ represented by arrows, 
since the second diagram has been added back. The cuts on the internal 
quark lines denote the insertion of the Fierz identity. 

The second diagram in Fig.~2(a) gives
\begin{eqnarray}
l_\beta\xi\not P_2\gamma^\beta\frac{1}{\xi\not P_2-\not l}
=\xi\not P_2(\not l-\xi\not P_2 +\xi\not P_2)\frac{1}{\xi\not P_2-\not l}
=-\xi\not P_2\;,
\label{ide2}
\end{eqnarray}
where the factor $\xi\not P_2$ at the beginning of the above expression 
comes from the internal quark propagator adjacent to the photon vertex, and
the term $-\xi\not P_2$ at the end  leads to the $O(\alpha_s^N)$ diagrams.
The diagrams $G_\parallel^{(N+1)}$ corresponding to the first term in 
Eq.~(\ref{dec}) are then factorized according to Fig.~2(b).  The factor 
$n_{+\alpha}/n_+\cdot l$ is just the Feynman rule associated with the 
Wilson line along the vector $n_+$ in Eq.~(\ref{pwj}). We arrive at the 
convolution for the first type of collinear configurations,
\begin{eqnarray}
G^{(N+1)}_{\parallel}&\approx&  G^{(N)}\otimes
J_\parallel^{(1)}\;,
\label{wicr}
\end{eqnarray}
with $J_\parallel^{(1)}$ given in Eq.~(\ref{j1c}).

The above procedures are applicable to the $O(\alpha_s^{j+1})$
jet function $J^{(j+1)}$. We identify the gluon in a complete set 
of $O(\alpha_s^{j+1})$ diagrams $J^{(j+1)}$, one of whose ends 
attaches the lower most vertex $\alpha$ on the internal quark line. The
other end attaches the vertex $\beta$ inside the rest of the diagrams. 
For the first term on the right-hand side of Eq.~(\ref{dec}),  there exists a 
Ward identity similar to Fig.~2(a). It is then trivial to obtain the 
factorization of the jet function corresponding to the first type of 
collinear configurations,
\begin{eqnarray}
J_{\parallel}^{(j+1)}&\approx& J^{(j)}\otimes
J_\parallel^{(1)}\;.
\label{wi2}
\end{eqnarray}

The rest part of the proof is exactly the same as of the factorization 
theorem for the semileptonic decay $B\to\pi l\bar\nu$ in Sec.~IV B
of \cite{NL}, with the pion distribution amplitudes replaced by the jet 
function, both of which are of the collinear origin. Employing 
Eqs.~(\ref{gnf}), (\ref{wicr}) and (\ref{wi2}) and following the 
steps of factorizing the decay $B\to\pi l\bar\nu$, we derive
\begin{eqnarray}
G^{(N+1)}=\sum_{m=+,-}
\sum_{i=0}^{N+1}\sum_{j=0}^{N+1-i}
\phi_m^{(i)}\otimes H_{m}^{(N+1-i-j)}\otimes J^{(j)}\;,
\label{gf2}
\end{eqnarray}
with the $O(\alpha_s^{N+1})$ hard amplitude $H_{m}^{(N+1)}$ being
infrared finite. Equation (\ref{gf2}) indicates that all the soft and
collinear divergences at the end point in the radiative decay 
$B\to\gamma l\bar\nu$ can be factorized into the $B$ meson distribution 
amplitudes and the jet function, respectively, order by order. 
We complete the proof of the factorization theorem, which is graphically 
described in Fig.~2(c), and construct the jet function defined 
in Eq.~(\ref{pwj}).

\section{JET FUNCTION}

We now perform threshold resummation of the double logarithms
$\alpha_s\ln^2 x$ in covariant gauge $\partial\cdot A=0$, which have been
collected into the jet function to all orders in the previous section. 
Threshold resummation for inclusive QCD processes has been studied 
intensively \cite{S0,CT}. Here we shall adopt the framework developed in 
\cite{L2}, which has been shown to lead to the same results as in 
\cite{S0,CT}. First, we replace the vector $n_+$ by $n$, which contains
a plus component $n^+>0$ and a (small) minus component. This 
replacement, regularizing the collinear pole, extracts the double 
logarithm as stated in Sec.~II. The jet functions in terms of the 
dimensionless vectors $n_+$ and $n$ can be regarded as being defined 
in the different factorization schemes. The definition then 
involves three variable vectors: the Wilson line direction $n$, the large 
momentum $P_2$, and the spectator momentum $k$. We argue that the 
jet function depends on the Lorentz invariants, $n\cdot k$, $n\cdot P_2$ 
and $P_2\cdot k$, since the other invariants $P_2^2$ and $k^2$ vanish, and  
$n^2$ will be fixed below. The third invariant $P_2\cdot k$ is not independent, 
because it can be rewritten as $n\cdot P_2 n\cdot k$ with $n^2$ being a 
constant. We further argue that the scale invariance in $n$, as indicated by the 
Feynman rule associated with the eikonal line along $n$, implies that
the jet function must depend on $k$ through the ratio $n\cdot k/n\cdot P_2$.

The next step is to derive the evolution of the jet function in $x$,
{\it i.e.}, in $k^+=xP_1^+$ by considering the derivative,
\begin{eqnarray}
k^+\frac{dJ}{dk^+}=\frac{n\cdot k}{P_2\cdot k}P_2^\alpha
\frac{dJ}{dn^\alpha}\;,
\end{eqnarray}
where we have applied the chain rule to relate the derivatives with respect to 
$k$ and to $n$. The differentiation $d/dn^\alpha$ operates 
on the eikonal line along $n$, giving
\begin{eqnarray}
\frac{n\cdot k}{P_2\cdot k}P_2^\alpha
\frac{d}{dn^\alpha}\frac{n_\mu}{n\cdot l}=\frac{{\hat n}_\mu}
{n\cdot l}\;,
\end{eqnarray}
with the special vertex,
\begin{eqnarray}
{\hat n}_\mu=-\frac{n\cdot k}{P_2\cdot k}\frac{P_2\cdot l}{n\cdot l}n_\mu\;.
\label{dp}
\end{eqnarray}
Another term in Eq.~(\ref{dp}), proportional to $P_2$, is negligible,
as the special vertex attaches inside of the jet function, which is
dominated by momenta parallel to $P_2$.

The loop momentum $l$ flowing through the special vertex does not
generate a collinear divergence due to vanishing of the numerator 
$P_2\cdot l$ in this region. It is easy to confirm that the ultraviolet 
region of $l$ does not produce $\ln x$ either. Therefore, we 
concentrate on the factorization of the soft gluon emitted from the special 
vertex, which can be achieved by applying the eikonal approximation to 
internal quark propagators, leading to $n_{-\nu}/n_-\cdot l$. Following the 
reasoning in \cite{L2}, the derivative of the jet function is written as
\begin{eqnarray}
x\frac{dJ(x)}{dx}=-ig^2C_F\int\frac{d^{4} l}
{(2\pi)^{4}}\frac{{\hat n}_\mu}{n\cdot l}
\frac{g^{\mu\nu}}{l^2}\frac{n_{-\nu}}{n_-\cdot l}J(x-l^+/P_1^+)\;,
\label{dij}
\end{eqnarray}
where the argument of $J$ in the integral arises from the invariant
mass of the internal quark,
$(P_2-k+l)^2\approx -2(x-l^+/P_1^+)P_1\cdot P_2$.
Performing the integration over $l^-$ and $l_T$, we derive the evolution
equation,
\begin{eqnarray}
x\frac{dJ(x)}{dx}=\frac{\alpha_s}{2\pi}C_F
\int_x^1\frac{d\xi}{\xi-x}J(\xi)\;,
\label{dj}
\end{eqnarray} 
with the integration variable $\xi$ corresponding to the plus component 
$l^+$. 

Employing the variable change $\xi=x/z$, Eq.~(\ref{dj}) becomes
\begin{eqnarray}
x\frac{dJ(x)}{dx}=J^{(0)}(x)+\frac{\alpha_s}{2\pi}C_F
\int_x^1\frac{dz}{z}J(x/z)\;,
\label{drj}
\end{eqnarray} 
where the nonperturbative inhomogenious term $J^{(0)}(x)$ absorbs 
the soft pole from $\xi\to x$ ($z\to 1$). After removing this pole, the integral
in the second term is infrared finite. It is interesting to observe that 
Eq.~(\ref{drj}) is similar to the evolution equation for unintegrated 
parton distribution functions involved in inclusive QCD processes \cite{KZ}, 
which resums the same double logarithm $\alpha_s\ln^2 x$. 

To solve Eq.~(\ref{dj}), we perform the Mellin transformation from the 
momentum  fraction ($x$) space to the moment ($N$) space:
\begin{eqnarray}
{\tilde J}(N)\equiv \int_0^1 dx (1-x)^{N-1} J(x)\;. 
\end{eqnarray}
The small $x$ region then corresponds to the large $N$ region.
The left-hand side of Eq.~(\ref{dj}) leads to
\begin{eqnarray}
\int_0^1 dx (1-x)^{N-1} x\frac{dJ(x)}{dx}=-N{\tilde J}(N)+(N-1){\tilde J}(N-1)
\approx-\frac{d}{dN}\left[N{\tilde J}(N)\right]\;,
\label{lh}
\end{eqnarray}
which is reasonable under the approximation $dN\approx \Delta N=1$. The 
right-hand side is given by
\begin{eqnarray}
\frac{\alpha_s}{2\pi}C_F\int_0^1  d\xi (1-\xi)^{N-1}J(\xi)
\int_0^1 \frac{dw}{1-w}\left(\frac{1-\xi w}{1-\xi}\right)^{N-1}
\approx\frac{\alpha_s}{2\pi}C_F\ln N{\tilde J}(N)\;,
\label{rh}
\end{eqnarray} 
where we have exchanged the sequence of the integrations over $x$ and
over $\xi$, made the variable change $x=\xi w$, and kept only the 
$N$-dependent term. The trick for extracting the $N$ dependence from the 
integral over $w$ in Eq.~(\ref{rh}) is to consider its derivative with respect to 
$N$, and to insert the large $N$ approximation for the $\delta$-function, 
$\lim_{N\to \infty}N(1-y)^{N-1}=\delta(y)$ with $y=\xi (1-w)/(1-\xi w)$.


Equating Eqs.~(\ref{lh}) and (\ref{rh}), we have the evolution
equation in the moment space,
\begin{eqnarray}
N\frac{d{\tilde J}(N)}{dN}=-{\tilde J}(N)-\frac{\alpha_s}{2\pi}C_F
\ln N {\tilde J}(N)\;,
\label{djn}
\end{eqnarray}
whose solution is the Sudakov factor,
\begin{eqnarray}
{\tilde J}(N)=\frac{1}{N}\exp\left(-\frac{1}{4}\gamma_K\ln^2 N\right)\;,
\label{j1}
\end{eqnarray}
with the anomalous dimension $\gamma_K=\alpha_sC_F/\pi$. The factor
$1/N$ can be regarded as the Mellin transformation of the initial condition
$J^{(0)}=1$. Comparing Eq.~(\ref{j1}) with the resummation for an inclusive 
jet \cite{S0,CT}, the resummation for an exclusive jet is suppressed by a 
factor $1/N$, and the corresponding anomalous dimension is down by a 
factor 2. 

By means of the inverse Mellin transformation, the jet function in the 
momentum fraction space is written as
\begin{eqnarray}
J(x)=\int_{c-i\infty}^{c+i\infty}\frac{dN}{2\pi i}(1-x)^{-N}
{\tilde J}(N)\;,
\label{mjn}
\end{eqnarray}
with $c$ being an arbitrary real constant larger than all the real parts of the 
poles of the integrand. Since  ${\tilde J}(N)$ contains a branch cut along the 
negative real axis on the complex $N$ plane, Eq.~(\ref{mjn}) reduces to
\begin{eqnarray}
J(x)=-\exp\left(\frac{\pi}{4}\alpha_sC_F\right)
\int_{-\infty}^{\infty}\frac{dt}{\pi}(1-x)^{\exp(t)}
\sin\left(\frac{1}{2}\alpha_s C_Ft\right)
\exp\left(-\frac{\alpha_s}{4\pi}C_Ft^2\right)\;,
\label{mjx}
\end{eqnarray}
where the variable change $N=\exp(t+i\pi)$ ($N=\exp(t-i\pi)$) has been
adopted for the piece of contour above (below) the branch cut as shown
in Fig.~3. Note that the above expression holds only for $\alpha_s>0$. 
For $\alpha_s=0$, the residue associated with the $N=0$ pole should be 
included. It is trivial to check that $J(x)$ is normalized to unity, 
$\int J(x)dx={\tilde J}(1)=1$ \cite{LL2}.

Obviously, Eq.~(\ref{mjx}) vanishes at $x\to 0$, because the integrand
is an odd function in $t$, and at $x\to 1$ due to the factor
$(1-x)^{\exp(t)}$. The latter property is the consequence of the
extrapolation of the Sudakov factor to the low $N$ region. Moreover, 
Eq.~(\ref{mjx}) provides suppression near the end point $x\to 0$, 
which is stronger than any power of $x$. This is understood from 
vanishing of all the derivatives of Eq.~(\ref{mjx}) with respect to $x$ 
at $x\to 0$. For example, the first derivative gives
\begin{eqnarray}
\frac{d}{dx}J(x)|_{x\to 0}
&=&-\exp\left(\frac{\pi}{4}\alpha_sC_F+\frac{\pi}{\alpha_sC_F} \right)
\int_{-\infty}^{\infty}\frac{dt}{\pi}e^t
\sin\left(\frac{1}{2}\alpha_s C_Ft\right)
\exp\left(-\frac{\alpha_s}{4\pi}C_Ft^2\right)\;,
\label{mj1}
\end{eqnarray}
where the variable change $t\to t+2\pi/(\alpha_sC_F)$ has been
made. The integrand in Eq.~(\ref{mj1}) is also an odd function in $t$, 
and the integral diminishes. To the accuracy of the next-to-leading 
logarithms, the running of the coupling constant $\alpha_s$ should 
be taken into account, and Eq.~(\ref{mjx}) will be modified. However, 
the above features remain.

\section{SEMILEPTONIC DECAY}

In this section we extend the above formalism to the semileptonic decay
$B\to\pi~l~{\bar\nu}$ in the fast recoil region of the pion. The $B$
meson momentum $P_1$ is the same as in the decay $B\to\gamma l\bar\nu$,
and the pion momentum $P_2$ is the same as the photon momentum.
Leading-twist factorization theorem for the $B\to\pi$ form factor $f_+(q^2)$ 
in the standard definition has been proved in \cite{L4}, 
\begin{eqnarray}
f_+(q^2)= \sum_{m=+,-}\phi_m(x_1)\otimes
H_m(x_1,x_2,\eta)\otimes \phi_\pi(x_2)\;,
\end{eqnarray}
which holds in the region with $x_1\sim O(\bar\Lambda/M_B)$ and 
with $x_2\sim O(1)$.

Since Fig.~4(a), proportional to $1/(x_1x_2^2)$, is more
singular at small $x_2$, we consider the end-point region with
$x_2\sim O(\bar\Lambda/M_B)$, where the internal $b$ quark propagator
scales like $1/({\bar\Lambda} M_B)$. Part of $O(\alpha_s)$ corrections to 
Fig.~4(a) are shown in Figs.~4(c)-(e), among which Fig.~4(c) generates the 
double logarithm $\alpha_s\ln^2 x_2$ from the collinear region with the
loop momentum parallel to $P_2$. Figures 4(d) and 4(e) involve only 
the single logarithm. We emphasize that the double logarithm discussed here 
for Fig.~4(a) differs from that in \cite{ASY}, which concerns $\alpha_s\ln^2 x_1$. 
As stated above, Fig.~4(a) is less singular at $x_1\to 0$, and 
$\alpha_s\ln^2 x_1$ is not relevant. The double logarithm in Fig.~4(c) 
is simply extracted by eikonalizing the light quark line, which flows into the 
pion, and the $b$ quark line according to Eq.~(\ref{ib}):
\begin{eqnarray}
J_{\parallel}^{(1)}&=&-ig^2C_F\int\frac{d^4 l}{(2\pi)^4}
\frac{n_+\cdot n_-}{n_+\cdot(l-x_2P_2)n_-\cdot l l^2}\;,
\nonumber\\
&=&-\frac{\alpha_s}{4\pi}C_F\ln^2 x_2+\cdots\;.
\label{j2c}
\end{eqnarray}
Similarly, the collinear pole has been regularized by including a small plus
component for $n_-$. It is observed that the double logarithm in 
Eq.~(\ref{j2c}) is the same as in Eq.~(\ref{j1c}).

The all-order factorization of the jet function from the decay 
$B\to\gamma l\bar\nu$ can be proved following the procedures in
Sec.~III.
It is also easy to show that this jet function obeys the evolution 
equation in Eq.~(\ref{dj}) by identifying the correspondence of
$x_2P_2$ and $n_+$ to $k$ and $P_2$ in Sec.~IV, respectively.
Hence, the threshold resummation leads to a result the same as 
Eq.~(\ref{mjx}). That is, the Sudakov factor is universal. 

The analysis for Fig.~4(b) is similar to that for the decay
$B\to\gamma l\bar\nu$. In the end-point region with
$x_1\sim O(\bar\Lambda^2/M_B^2)$, additional collinear divergences
associated with the internal light quark are produced in Figs.~4(f)-4(h).
Figure 4(f) gives the double logarithm $\alpha_s\ln^2 x_1$, whose 
factorization is the same as of Fig.~1(c). Note that
Fig.~5 has been considered as the dominant source of the double 
logarithm $\alpha_s\ln^2 x_1$ associated with Fig.~4(b) \cite{ASY}. 
It will be demonstrated explicitly in the Appendix A that Fig.~5 
in fact contains only the single logarithm $\alpha_s\ln x_1$. The 
result in \cite{ASY} might be attributed to an inappropriate 
collinear approximation. Therefore, the universality of the 
double-logarithm resummation indeed holds.

At last, we investigate the resummation effect on the $B\to\pi$ form
factor $f_+(q^2)$, whose factorization formula is given by
\begin{eqnarray}
f_+(q^2)&=&\frac{\pi \alpha_s C_Ff_Bf_\pi}{\eta^2 M_B^2 N_c}\int dx_1dx_2
\phi_B(x_1)\left[\frac{J(x_2)}{x_1x_2^2}(1+x_2\eta)
-\frac{J(x_1)}{x_1x_2}(1-\eta)\right]\phi_\pi(x_2)\;,
\label{f+}
\end{eqnarray}
with $N_c=3$ being the number of colors. We employ the models
\cite{PB1,BW},
\begin{eqnarray}
\phi_\pi(x)&=&6x(1-x)\{1+0.66[5(1-2x)^2-1]\}\;,
\nonumber\\
\phi_B(x)&\equiv&\frac{1}{2}[\phi_+(x)-\phi_-(x)]
=N_B\sqrt{x(1-x)}
\exp\left[-\frac{1}{2}\left(\frac{xM_B}{\omega_B}\right)^2\right]\;,
\label{os}
\end{eqnarray}
with the shape parameter $\omega_B$ and the normalization constant $N_B$
determined by $\int \phi_B(x)dx=1$. Another $B$ meson distribution amplitude 
${\bar\phi}_B(x)=[\phi_+(x)-\phi_-(x)]/2$ is power-suppressed, since its
first nonvanishing moment starts with $O(1/M_B)$ \cite{TLS}. It has been
confirmed, contrary to the conclusion in \cite{GS}, that the contribution from 
${\bar\phi}_B$ is less important than that from $\phi_B$ \cite{WY}. For 
the purpose of this work, it is sufficient to work on a single $B$ meson 
distribution amplitude. If $J(x)$ is excluded, the first
term in Eq.~(\ref{f+}) is logarithmically divergent. With the threshold
resummation, the form factor is calculable without introducing any infrared
cutoffs \cite{SHB,ASY,BF}.

Choosing $\alpha_s=0.4$, $M_B=5.28$ GeV, and the decay constants
$f_B=190$ MeV and $f_\pi=130$ MeV, we obtain $f_+(q^2)$
for $\omega_B=0.3(0.4)$ GeV,
\begin{eqnarray}
f_+(0)=0.15(0.12)\;,\;\;\;
f_+(2\;\;{\rm GeV}^2)=0.18(0.13)\;,\;\;\;
f_+(4\;\;{\rm GeV}^2)=0.21(0.15)\;.
\end{eqnarray}
The variation of $f_+$ for $\alpha_s=0.3$-0.5 is less than 10\%, because
the change of $J$ is compensated by that of $\alpha_s$ in the overall
coefficient in Eq.~(\ref{f+}). The difference between the above values
and the expected one $f_+(0)\sim 0.3$ can be resolved by taking into
account higher-twsit contributions \cite{TLS}. Including two-parton twist-3 
pion distribution amplitudes, which are finite at the end point \cite{PB1}, 
the singularities in the $B\to\pi$ form factor become linear. In this 
case threshold resummation is even more crucial.

\section{CONCLUSION}

In this paper we have shown that the double logarithmic corrections 
$\alpha_s\ln^2 x$ appear in exclusive $B$ meson decays. When the 
end-point region with a momentum fraction $x\to 0$ is important, these 
double logarithms need to be organized to all orders in order to justify 
perturbative expansion. The double logarithms, associated with
internal particles which almost go on mass shell, are of the collinear 
nature. The PQCD factorization theorem for exclusive $B$ 
meson decays at the end point then demands the introduction of a quark 
jet function into decay amplitudes. The factorization of the jet function 
from the $B\to\gamma l\bar\nu$ mode has been proved rigorously. The 
proof for the factorization from other modes, such as the semileptonic decay
$B\to\pi l\bar\nu$, is similar. It has been shown that the jet function is 
defined as a matrix element of a quark field attached by a Wilson line,
based on which threshold resummaiton can be performed. It is
interesting to achieve the above proof in the framework
of the soft-collinear effective theory \cite{BPS}.

We have derived the evolution equation for the jet function, which is
solvable in the large $N$ limit in the Mellin space. The solution is a 
universal, {\it i.e.}, process-independent, Sudakov factor. 
The qualitative behavior of this Sudakov factor has analyzed and 
found to decrease quickly at $x\to 0$. It has been demonstrated that
Sudakov suppression is strong enough to smear the end-point 
singularity in the $B\to\pi$ form factor and leads to reasonable predictions. 
We conclude that in a self-consistent PQCD analysis of the 
heavy-to-light transition form factors, the end-point singularities do not 
exist. In a future work we shall extend the formalism developed here to
more complicated nonleptonic $B$ meson decays. Threshold resummation
for various topologies of decay amplitudes, such as annihilation and 
nonfactorizable ones, will be studied.  

\vskip 0.3cm
I thank M. Beneke, Y.Y. Keum, T. Kurimoto, D. Pirjol, A.I. Sanda, and
G. Sterman for useful discussions. The work was supported in part by the
National Science Council of R.O.C. under Grant No. NSC-90-2112-M-001-077,
by the National Center for Theoretical Sciences of R.O.C.,
by Grant-in Aid for Special Project Research (Physics of CP violation),
and by Grant-in Aid for Scientific Exchange from the Ministry of
Education, Science and Culture of Japan.

\appendix

\section{TRIPLE-GLUON CORRECTION}

In this Appendix we show that Fig.~5 produces only the single logarithm
$\alpha_s\ln x_1$. The loop integral is written as
\begin{eqnarray}
I&=& g^4\int\frac{d^4 l}{(2\pi)^4}
tr\left[\gamma_\delta\frac{\gamma_5\not P_2}{4N_c}\gamma_\beta
\frac{\not P_2-\not k_1 +\not l}{(P_2-k_1+l)^2}\gamma^\mu\right.
\nonumber\\
& &\times \left.\frac{\not P_1-\not k_1 +\not l+M_B}{(P_1-k_1+l)^2-M_B^2}
\gamma_\alpha\frac{(\not P_1+M_B)\gamma_5}{4N_c}\right]
\nonumber\\
& &\times \frac{\Gamma^{\alpha\beta\delta}f^{abc}tr(T^aT^bT^c)}
{(k_1-x_2P_2-l)^2 l^2(x_2P_2-k_1)^2}\;,
\label{i1}
\end{eqnarray}
with the triple-gluon vertex,
\begin{eqnarray}
\Gamma^{\alpha\beta\delta}=(2x_2P_2-2k_1+l)^\alpha g^{\beta\delta}
+(k_1-x_2P_2-2l)^\delta g^{\alpha\beta}+(l-x_2P_2+k_1)^\beta
g^{\alpha\delta}\;,
\label{tri}
\end{eqnarray}
the anti-symmetric tensor $f^{abc}$, and the color matrices $T^{a,b,c}$.

We concentrate on the frist term in Eq.~(\ref{tri}), since the calculation
of the second and third terms is similar. If the loop momentum $l$ in
the quark propogator $1/(P_2-k_1+l)^2$ is neglected, Eq.~(\ref{i1})
gives the double logarithm $\alpha_s\ln^2 x_1$ as obtained in
\cite{ASY}. However, this approximation is not
appropriate for the collection of the collinear divergence, for which
$l$ is of the same order as $P_2$. Keeping the $l$ dependence, and using
the identity $f^{abc}tr(T^aT^bT^c)=6i$, we have
\begin{eqnarray}
I_1&=& -\frac{ig^4}{4N_c} \frac{tr(\not P_2\not k_1 \gamma^\mu\not P_1)}
{(x_2P_2-k_1)^2}
\nonumber\\
& &\times\int\frac{d^4 l}{(2\pi)^4}\frac{v\cdot (2x_2P_2-2k_1+l)}
{v\cdot l(P_2-k_1+l)^2(k_1-x_2P_2-l)^2 l^2}\;.
\label{i2}
\end{eqnarray}
The contribution proportional to $M_B^2$ in Eq.~(\ref{i1}) is suppressed
by a power of $x_1$ compared to Eq.~(\ref{i2}). For a similar reason,
the term $v\cdot k_1$ in Eq.~(\ref{i2}) is also negligible.
The term $v\cdot l$ in the numerator does not
lead to a collinear divergence, because the poles of the corresponding
denominator $(P_2-k_1+l)^2(k_1-x_2P_2-l)^2 l^2$ in the $l^-$ plane are 
not of the pinched type.

Doing the contour integral with the residue $l^-=-l^+-i\epsilon$
(for $l^+ < 0$), Eq.~(\ref{i2}) becomes
\begin{eqnarray}
I_1&=& H^{(0)\mu}_{4b}\frac{4g^2}{(2\pi)^3C_F}
\int_{-\infty}^0 dl^+ \int_{-\infty}^{\infty} d^2l_T
\nonumber\\
& &\times
\frac{x_2 P_2^-(P_2-k_1)^2}
{(l_T^2+2l^{+2}-2P_2^-l^+ +2P_2^-k_1^+)
(l_T^2+2l^{+2}-2x_2P_2^-l^+ +2x_2P_2^- k_1^+)
(l_T^2+2l^{+2})}\;,
\label{i3}
\end{eqnarray}
where
\begin{eqnarray}
H^{(0)\mu}_{4b}=\frac{g^2C_Ftr(\not P_2\not k_1 \gamma^\mu\not P_1)}
{8N_c(x_2P_2-k_1)^2(P_2-k_1)^2}\;,
\end{eqnarray}
is the lowest-order amplitude from Fig.~4(b). The integration over $l_T$ 
leads to
\begin{eqnarray}
I_1&=&H^{(0)\mu}_{4b}\frac{\alpha_s}{4\pi}C_A x_2k_1^+
\int_0^{\infty} dl^+\left[\frac{\ln(l^{+2}/M_B^2
+\eta l^+/M_B +x_1\eta)}{(1-x_2)(l^+ +k_1^+)^2}\right.
\nonumber\\
& &\left.-\frac{\ln(l^{+2}/M_B^2+x_2\eta l^+/M_B +x_1x_2\eta)}
{x_2(1-x_2)(l^+ +k_1^+)^2}
+\frac{\ln(l^{+2}/M_B^2)}{x_2(l^+ +k_1^+)^2}\right]\;,
\label{i4}
\end{eqnarray}
with the color factor $C_A=3$.

The terms $l^{+2}/M_B^2$ in the first two logarithms, giving contributions
suppressed by a power of $x_1$ compared to those from $l^+/M_B$, can be
dropped. Each term in Eq.~(\ref{i4}) then
produces only the single logarithms $\alpha_s\ln x_1$:
\begin{eqnarray}
I_1&=&H^{(0)\mu}_{4b}\frac{\alpha_s}{4\pi}C_A x_2
\left[\frac{\ln x_1}{1-x_2}-\frac{\ln x_1}{x_2(1-x_2)}
+2\frac{\ln x_1}{x_2}\right]\;,
\nonumber\\
&=&H^{(0)\mu}_{4b}\frac{\alpha_s}{4\pi}C_A \ln x_1\;.
\label{i5}
\end{eqnarray}
In the above expression the constant terms independent of
$\ln x_1$ are not displayed. It can be shown in a similar way that the 
other two terms in Eq.~(\ref{tri}) involve only the single logarithms.


\vskip 0.5cm
\centerline{\large \bf Figure Captions}
\vskip 0.3cm

\noindent
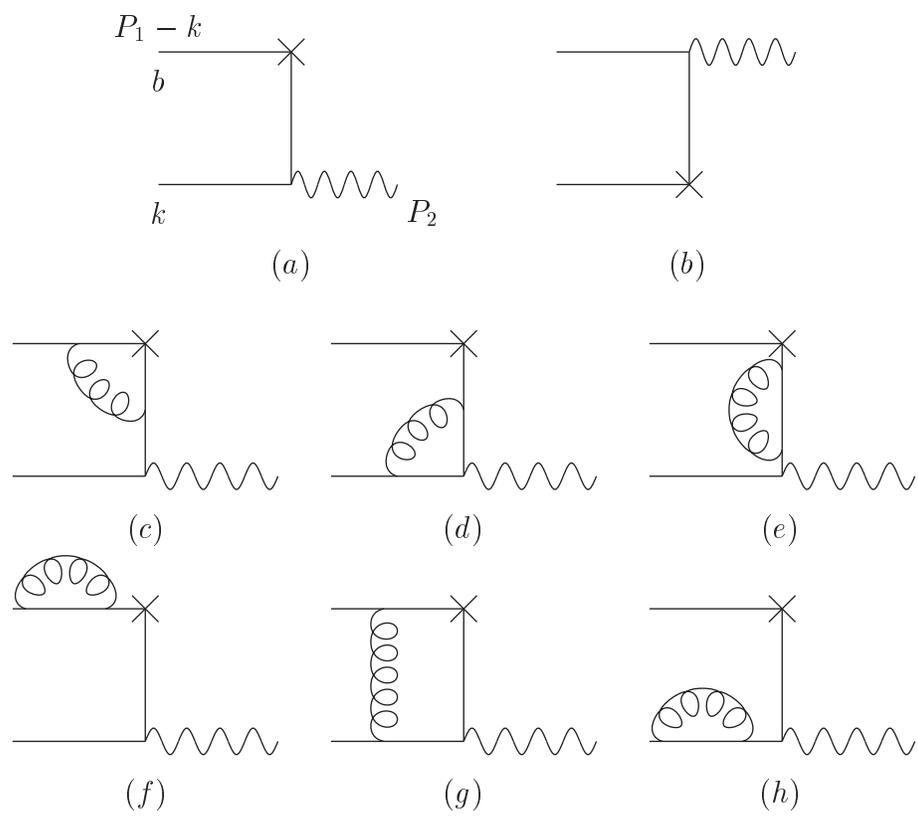
{\bf Fig. 1.} (a) and (b) Lowest-order diagrams for the decay
$B\to\gamma l\bar\nu$. (c)-(h) $O(\alpha_s)$ corrections to Fig.~1(a).
\vskip 0.3cm

\noindent
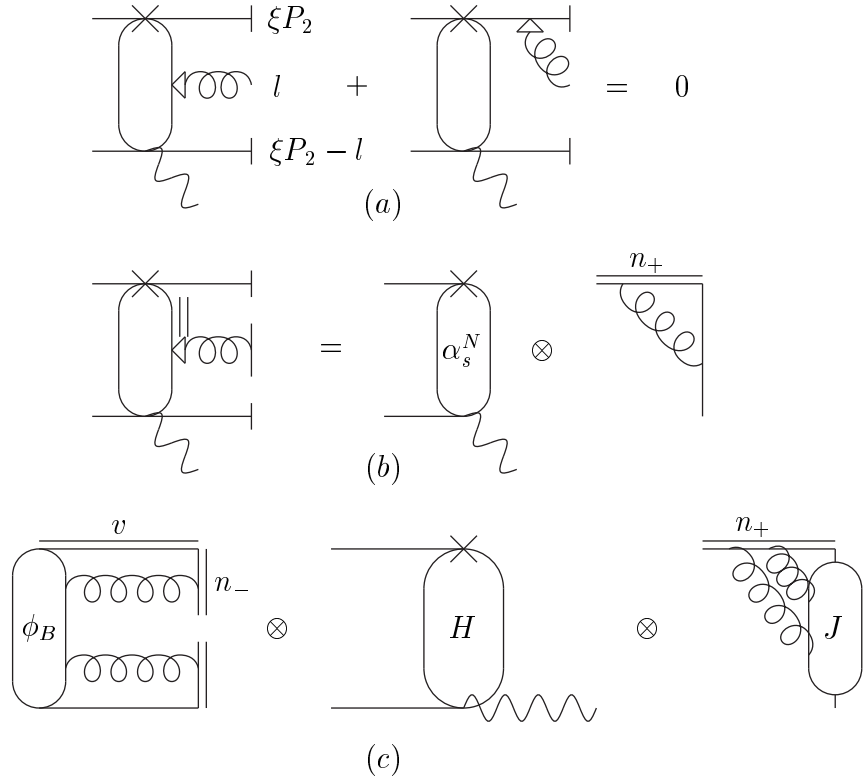
{\bf Fig. 2.} (a) The Ward identity. (b) Factorization
of the $O(\alpha_s)$ diagrams as a consequence of (a).
(c) Factorization of the $B\to\gamma l\bar\nu$ decay amplitude in the 
end-point region.
\vskip 0.3cm 

\noindent
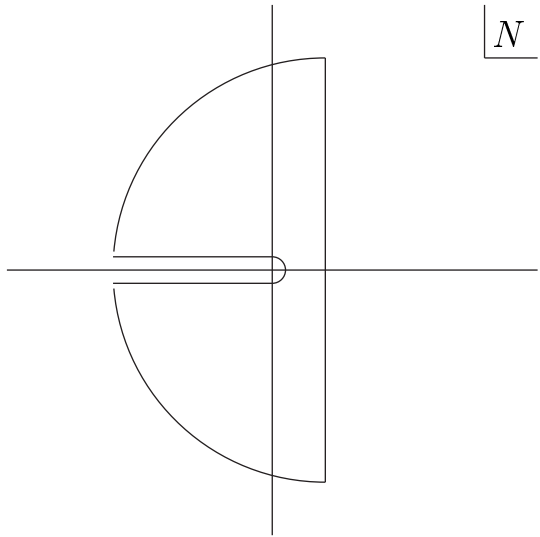
{\bf Fig. 3.} Contour for the inverse Mellin transformation of the
jet function.
\vskip 0.3cm 

\noindent
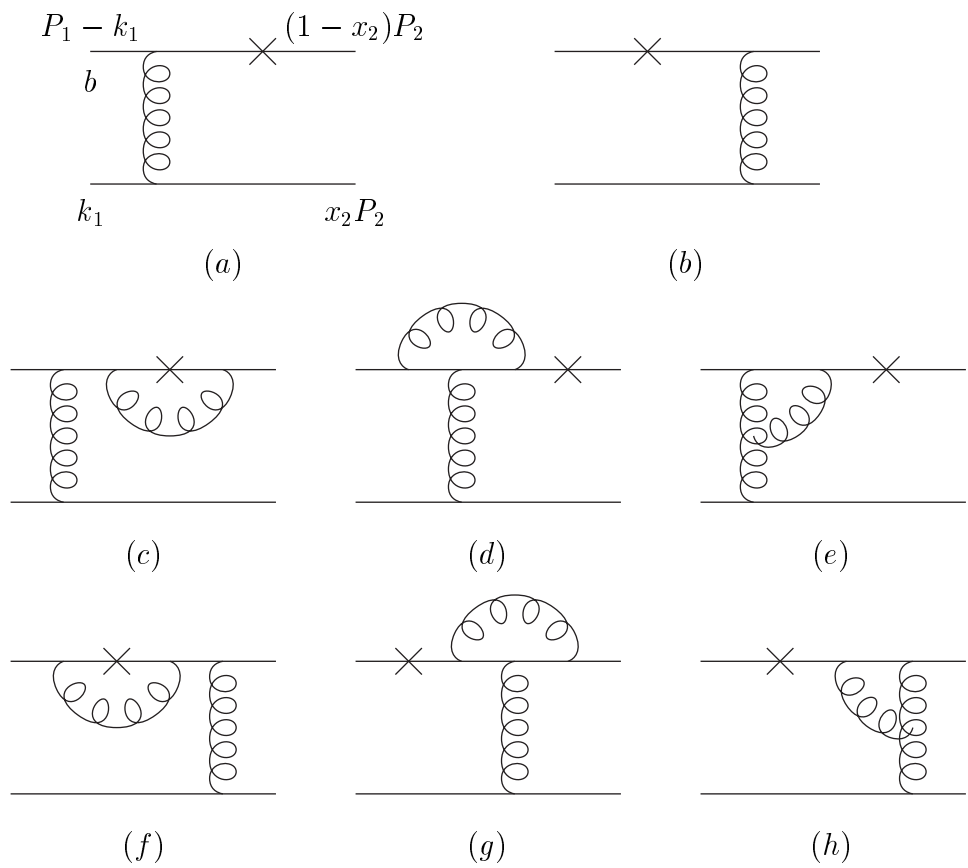
{\bf Fig. 4.} (a) and (b) Lowest-order diagrams for the decay
$B\to\pi l\bar\nu$. (c)-(e) [(f)-(h)] Part of $O(\alpha_s)$ corrections to
Fig.~4(a) [Fig.~4(b)].
\vskip 0.3cm

\noindent
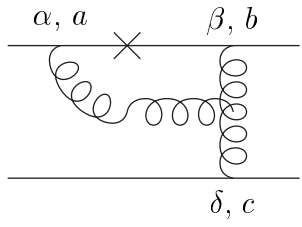
{\bf Fig. 5.} Triple-gluon correction to Fig.~4(b).

\end{document}